\documentclass[times,singlecolumn,final]{elsarticle} %这边改变单列还是双列
\usepackage{medima}
\usepackage{framed,multirow}

\usepackage{amsmath,amssymb,amsfonts}
\usepackage{latexsym}

\usepackage{url}
\usepackage{xcolor}

\usepackage{hyperref}
\usepackage{bm}
\usepackage{color}
\usepackage{booktabs}
\usepackage{threeparttable}
\definecolor{newcolor}{rgb}{.8,.349,.1}

\journal{Medical Image Analysis}

\begin{document}

\verso{K.Wang \textit{et~al.}}

\begin{frontmatter}

\title{AWSnet: An Auto-weighted Supervision Attention Network for Myocardial Scar and Edema Segmentation in Multi-sequence Cardiac Magnetic Resonance Images}%
%%\tnotetext[tnote1]{This is an example for title footnote coding.}

\author[1]{Kai-Ni Wang \fnref{fn1}}
%%\fntext[fn1]{This is author footnote for second author.}
\author[2]{Xin Yang \fnref{fn1}}
\fntext[fn1]{Kai-Ni Wang and Xin Yang contribute equally to this work.}
\author[1]{Juzheng Miao }
\author[3,4,5]{Lei Li }
%% Third author's email
%%\ead{author3@author.com}
\author[6]{Jing Yao }
\author[1]{Ping Zhou}
\author[2]{Wufeng Xue}
\author[1]{Guang-Quan Zhou \corref{cor1}}
\author[3]{Xiahai Zhuang \corref{cor1}}
\author[2]{Dong Ni \corref{cor1}}
\cortext[cor1]{Corresponding author:
	    e-mail address: guangquan.zhou@seu.edu.cn (Guang-Quan Zhou), zxh@fudan.edu.cn (Xiahai Zhuang), nidong@szu.edu.cn (Dong Ni)
}

\address[1]{State Key Laboratory of Bioelectronics, School of Biological Science and Medical Engineering, Southeast University, Nanjing, China}
\address[2]{Medical UltraSound Image Computing (MUSIC) Lab, School of Biomedical Engineering, Health Center, Shenzhen University, Shenzhen, China}
\address[3]{School of Data Science, Fudan University, Shanghai, China}
\address[4]{School of Biomedical Engineering, Shanghai Jiao Tong University, Shanghai, China}
\address[5]{School of Biomedical Engineering and Imaging Sciences, Kings College London, London, UK}
\address[6]{Department of Ultrasound Medicine, Affiliated Drum Tower Hospital of Nanjing University Medical School, Nanjing, China}

\received{1 May 2013}
\finalform{10 May 2013}
\accepted{13 May 2013}
\availableonline{15 May 2013}
\communicated{S. Sarkar}

\begin{abstract}
%%%
Multi-sequence cardiac magnetic resonance (CMR) provides essential pathology information (scar and edema) to diagnose myocardial infarction. However, automatic pathology segmentation can be challenging due to the difficulty of effectively exploring the underlying information from the multi-sequence CMR data. This paper aims to tackle the scar and edema segmentation from multi-sequence CMR with a novel auto-weighted supervision framework, where the interactions among different supervised layers are explored under a task-specific objective using reinforcement learning. Furthermore, we design a coarse-to-fine framework to boost the small myocardial pathology region segmentation with shape prior knowledge. The coarse segmentation model identifies the left ventricle myocardial structure as a shape prior, while the fine segmentation model integrates a pixel-wise attention strategy with an auto-weighted supervision model to learn and extract salient pathological structures from the multi-sequence CMR data. Extensive experimental results on a publicly available dataset from Myocardial pathology segmentation combining multi-sequence CMR (MyoPS 2020) demonstrate our method can achieve promising performance compared with other state-of-the-art methods. Our method is promising in advancing the myocardial pathology assessment on multi-sequence CMR data. To motivate the community, we have made our code publicly available via \href{https://github.com/soleilssss/AWSnet/tree/master}{ https://github.com/soleilssss/AWSnet/tree/master}.
%%%%
\end{abstract}


\end{frontmatter}

%\linenumbers

%% main text
\begin{figure*}[!t]
	\centering
	\includegraphics[width=\textwidth]{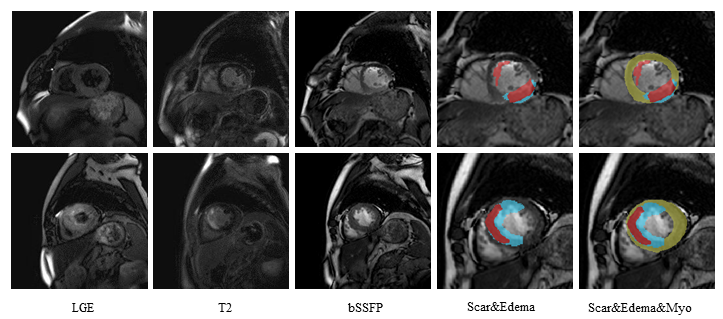}
	\caption{Visualization of LV myocardial scar and edema on three sequences of CMR images, including LGE, T2, and bSSFP. The fourth and fifth column is the bSSFP CMR image overlapped with annotations, where the blue area indicates edema, and the red area indicates scars. LV myocardium is indicated in yellow in the fifth column.}
	\label{img1}
\end{figure*}

\section{Introduction}
\label{sec1}
Myocardial infarction (MI) is a common cardiovascular disease affecting millions of patients globally, which has become one of the leading causes of death worldwide
\citep{mcnamara2016predicting}.
Early accurate quantification analysis of infarcted myocardium (Myo) is very important for effective diagnosis and therapy to mitigate MI mortality risk
\citep{white2008acute}.
The infarct region extent is also highly correlated with the prognosis of MI 
\citep{roes2007comparison}.
Medical imaging plays a crucial role in assessing MI by measuring myocardial scar and edema that are critical clinical presentations to reflect MI severity
\citep{richardson2011physiological}.In particular, cardiac magnetic resonance (CMR), the gold-standard technique for noninvasive myocardial tissue characterization, can provide comprehensive anatomical and functional information about the myocardium  
\citep{ibanez2019cardiac}. The delineation of the healthy and pathological myocardium borders, known as segmentation, is mandatory and critical in clinical CMR analysis. Recent studies have revealed that the knowledge of infarct size, shape, and location can help determine the correct ablation locations and size
\citep{estner2011critical,andreu2011integration}. However, the manual delineation of pathological regions from the myocardium is generally time-consuming and potentially subjective. Therefore, automatic scar and edema segmentation from CMR images are of great clinical significance to quantify cardiac morphological and pathological changes, facilitating treatment planning and patient management.

Single-sequence late gadolinium enhancement (LGE) CMR has been commonly employed to visualize the myocardial pathology with the intensity distribution and subsequently quantitatively analyze the scar and edema \citep{kim2009cardiovascular,flett2011evaluation}. The LGE CMR sequence presents the infarcted myocardium with distinctive brightness than the healthy tissues. Several studies automatically separate the scar and edema from the normal myocardium using conventional intensity-based methods, such as the Gaussian mixture model (GMM)-based segmentation
\citep{balafar2014gaussian}, the signal threshold to reference mean (STRM)
\citep{kolipaka2005segmentation}, and region growing (RG)
\citep{alba2012healthy}. Moreover, multi-sequence CMR can provide rich and reliable information about the pathology and morphology of the myocardium. For example, the LGE sequence can highlight the pathological areas, T2-weighted CMR can depict edema
\citep{kim2009cardiovascular}, and the balanced-Steady State Free Precession (bSSFP) cine sequence captures cardiac motions and presents clear boundaries \citep{flett2011evaluation}.
Fig.\ref{img1} presents an example of three CMR sequences, including LGE, T2, and bSSFP. However, it is still hard to combine multi-sequence CMR data effectively in the myocardial pathological segmentation. There are three major challenges related to this myocardial pathological segmentation. First, the heterogeneous characterization of infarcted regions across different patients makes it difficult to construct the prior model of scar and edema. Moreover, the scar and edema regions distributed at various locations occupy tiny area on the myocardium, greatly increasing the difficulty of automatic segmentation. The boundary between scar and edema is often indistinct in the CMR images. Finally, different CMR sequence enhancement patterns can be complicated. 

To our knowledge, the literature related to myocardial pathological segmentation hardly contemplates these differences among these multi-source images. This paper, therefore, propose a novel coarse-to-fine framework with auto-weighted supervision and pixel-wise attention submodules, namely AWSnet, to segment the myocardial pathology (scar and edema) from multi-sequence CMR images by exploring the prior information of the left ventricle (LV). Our major contributions are summarized as follows:
\begin{enumerate}
	\item Our proposed two-stage framework exploits prior knowledge and anatomical dependence of scar and edema to boost the small myocardial pathology region segmentation. The coarse segmentation model identifies the LV myocardial structure as the guide to the scar and edema segmentation.  The second-stage attention-based U-Net drops the bSSFP CMR to extract salient pathological structures from LGE and T2-weighted CMR sequences. Experiments on a public dataset demonstrated the effect of the proposed AWSnet, which achieved competitive performance over the state-of-the-art. In addition, we verified the generalization performance of our method on the ACDC and BUSI datasets.
	\item This study deems the weighting of supervision layers as the model hyperparameters to tune the contributions from different supervision layers adaptively. To the best of our knowledge, our auto-weighted method is the first that leverages the reinforcement learning (RL) based method to dynamically explore the interactions among different supervised layers under a task-specific objective. This RL-based method effectively learns the optimal weighting of different supervision layers, thereby achieving better performance than traditional deep supervision methods.
	\item The pixel-wise attention module enables the fine segmentation network to scrutinize the latent pathological structure even with large shape variability, improving the scar and edema detection at a pixel level.
\end{enumerate}

The rest of this paper is organized as follows. The next section
reviews the related works. A detailed explanation of our proposed AWSnet method is described in section \ref{sec3}. Section
\ref{sec4} and 
\ref{sec5} present experimental results and corresponding analysis. Finally, section
\ref{sec6} concludes the proposed work.

\section{Related Works}
\subsection{Myocardial Pathological Segmentation}
Limited studies have been reported in the literature to develop automated myocardial scar and edema segmentation. Most of such work relies on a prerequisite to integrate the prior shape information of the myocardium. Several reported methods employ thresholding segmentation based on the image intensity probability distributions among healthy and pathological tissues 
\citep{kolipaka2005segmentation,alba2012healthy}. However, these thresholding methods are vulnerable to image noises
\citep{gao2013highly,zhang2016myocardial}.
\citep{ukwatta2015myocardial}
 depicts the infarct segmentation as a continuous max-flow (CMF) optimization problem, in which manual segmentation of the LV myocardium is used to initialize and constrain the method. In 
\citep{merino2016variational}, a variational framework is employed to restrain the myocardium shape to identify healthy and scar tissues simultaneously. 

The deep neural networks (DNN) have revolutionized various medical image segmentation tasks by seamlessly integrating mid- and high- levels of image features 
\citep{2018Deep,ge2019pv,huang2020segmentation}. For pathology segmentation tasks on the myocardium, most of the studies presented are based on the U-Net
 architecture, a classical encoder-decoder segmentation network 
\citep{falk2019u}.
\citep{chen2018multiview} employ a multiview recursive attention model to segment the left atrium (LA) myocardium and pulmonary veins (PV) anatomy and the atrial scars sequentially from the LGE-MRI images.
\citep{li2020joint} design a modified U-Net model to segment LA and quantify the scars amount simultaneously, where the scar quantification was performed on the LA boundary. Meanwhile, they propose a scar segmentation method based on the graph-cuts framework, where multi-scale learning combined with DNN is used to estimate the edge weights of a graph
\citep{li2020atrial}.
\citep{zabihollahy2019convolutional}
  introduce a fully convolutional network (FCN) to segment scar from the manual delineation of the LV myocardium region in LGE CMR images. They further advance a multi-planar U-Net network to segment LV myocardium to realize a fully automatic scar segmentation
\citep{zabihollahy2020fully}.

Most DNN-based myocardial pathology segmentation only focuses on mono-sequence CMR, such as LGE. The images acquired from different sequence CMR can be deemed as a separated modality, providing different meaningful information of the whole heart. Many learning-based approaches have been proposed to fuse multi-modality information, substantially improving learning efficiency and prediction accuracy for different tasks
\citep{baltruvsaitis2018multimodal}. The fusion strategy is task-dependent though the most widely used fusion techniques are data integration (early fusion) and decision integration (late fusion) 
\citep{vielzeuf2018centralnet}. Recently, 
\citep{zhang2020multi}  propose a multi-modality network for scar segmentation, which employs the output of an anatomical structure segmentation network to regulate the following pathological region segmentation model equipped with a channel attention-based fusion block. However, these methods still do not deliberately consider the differences among bSSFP cine CMR and LGE and T2-weighted CMR sequences. They also neglect the small size and fuzzy border in the edema and scar segmentation. In this paper, given multi-sequence CMR images, namely bSSFP, T2, LGE, we design a novel coarse-to-fine segmentation framework to fully exploit the prior information on the modality difference and anatomical dependence for automatic scar and edema segmentation. Moreover, we fuse a reinforcement learning-based deep supervision model into the pixel-wise attention network to better focus on salient pathological structures segmentation.

\subsection{Deep Supervision}
Deep supervision, firstly introduced by
\citep{lee2015deeply}, has been demonstrated as an effective practice to improve network performance, such as in medical image segmentation 
\citep{dou20173d} and object detection 
\citep{shen2019object}. Its key point is to supervise the earlier hidden layers as companion objective functions rather than the objective function only at the output one. These auxiliary objective functions from intermediate layers can mitigate gradient vanishing problems during training and speed up convergence. Furthermore, it can act as a feature regularization method on the overall model performance, boosting semantically meaningful deep features. From another view, the companion objective functions at each hidden layer can be deemed as adding additional tasks at multiple resolutions to the network. Deep supervision usually follows a heuristic approach to predefine initial weights of companion supervision losses corresponding intermediate layers and decay them during training. However, there is an inherited inconsistency and uncertainty among different tasks. It is desirable to emphasize one or two supervision layers from all the available information during learning, like multi-modality learning. We can hardly identify the more essential layers for different tasks in advance by defining these hyperparameters heuristically. Recently, automating meta-parameters has become the most promising research field to tackle knowledge versatility for various computer vision tasks, such as image recognition \citep{zoph2018learning, liu2018progressive}. It is highly desired to learning the weighting of different supervision layers as the model hyperparameters to optimize the interactions among different supervised layers under a task-specific objective for medical image segmentation. In this work, we empower deep supervision with an elegant and implicit solution, called deep auto-weighted supervision, to dynamically optimize the balancing weight of supervision loss using an RL framework, beneficial to emphasize task-specific semantic representations to the intermediate layers.

\subsection{Deep Reinforement Learning}
Reinforcement learning (RL) is defined as a computational approach to reach predefined goals by interacting with an environment maximizing cumulative reward signals 
\citep{sutton2018reinforcement}. The RL has shown great potential by integrates the powerful understanding ability of DNN in decision-making to achieve accurate detection and segmentation results. Some RL-based methods have directly explored the landmark location by learning the optimal path to the target with the maximum accumulated rewards of taking sequential action steps with a multi-scale strategy \citep{ghesu2017multi,alansary2019evaluating} or a dedicated reward design \citep{zhang2020enhanced}. Recently, 
\citep{zhang2020sequential} model the spatial correlation of different vertebral bodies using sequential dynamic-interaction processes and propose a sequential conditional RL network to achieve the detection and segmentation of the vertebral body from MR spine images simultaneously. On the other hand, RL becomes a prospective research field in automating meta-parameters optimization, such as network architecture search and objective metrics learning. In 
\citep{zoph2016neural}, the network architecture is iteratively searched to optimize the classification performance using RL. Under a similar RL framework, 
\citep{cubuk2019autoaugment} automatically search the optimal augmentation policies that the neural network yields the highest validation accuracy on a target dataset. Meanwhile, 
\citep{li2019lfs} leverage the RL approach to search loss functions from the search space in which some existing prevailing loss functions are integrated into a unified formulation. However, to the best of our knowledge, little work has been conducted to determine the task-specified learning strategy rationally, especially in deep supervision learning. In this study, the RL methods with strong decision-making ability highly inspire us to exploit RL to probe the relationship among the supervised layers in the deep supervision network.

\begin{figure*}[!t]
	\centering
	\includegraphics[width=\textwidth]{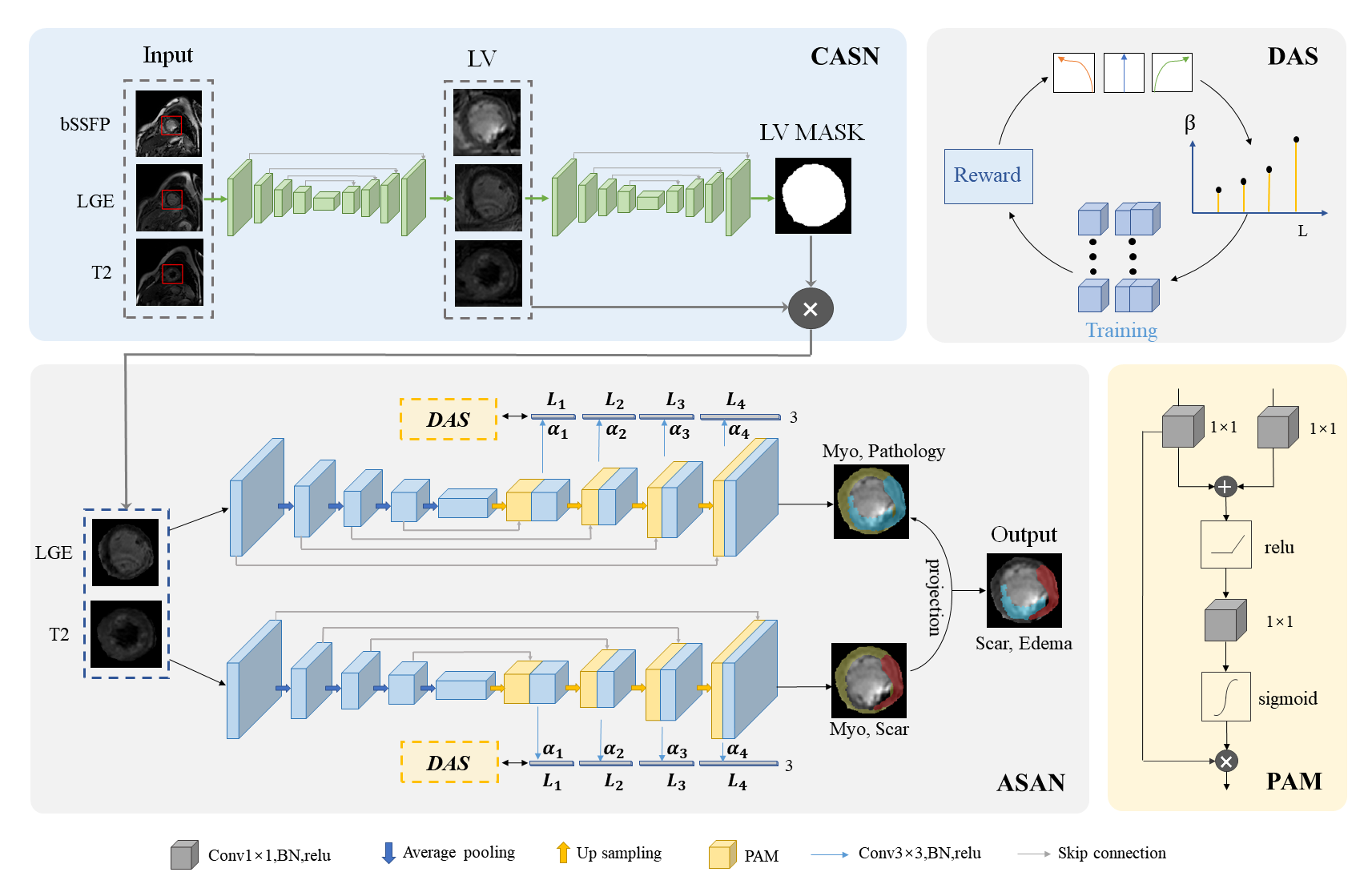}
	\caption{The flow diagram of the proposed coarse-to-fine framework with an auto-weighted supervision model. In the coarse segmentation stage, all three CMR sequences are fed to a cascaded vanilla U-Net for LV segmentation. The regions enclosed by the red rectangles are cropped to the second vanilla U-Net. In the following fine segmentation stage, the bSSFP CMR sequence is neglected from scratch in the pixel-wise attention of U-Net to identify scar and edema effectively. The optimal weighting       $\bm{\left(\alpha_{1},\alpha_{2}, \alpha_{3},\alpha_{4}\right)}$ among supervised layers is obtained by the RL-based auto-weighted module (DAS).}
	\label{zongkuangtu}
\end{figure*}
\section{Methods}
\label{sec3}
Fig.\ref{zongkuangtu} presents the flow diagram of our proposed AWSnet, which integrates deep auto-weighted supervision and pixel-wise attention for the myocardial pathology segmentation from the multi-sequence CMR data with an end-to-end framework:

\textbf{Two-stage framework} segments myocardial pathology with a coarse-to-fine approach. In the first stage, we integrate multi-sequences CMR images, namely bSSFP, LGE, and T2, in an early-fusion manner to input cascaded anatomical segmentation network (CASN), contributing to data-balance and the prior of spatial dependence in the following myocardial pathology segmentation. Subsequently, the LGE and T2 CMR data in the guidance of the segmented LV structure are concatenated to the deep auto-weighted supervision attention network (ASAN) for the scar and edema segmentation. Finally, we apply ensemble learning to improve the myocardial pathology segmentation performance.

\textbf{Deep auto-weighted supervision (DAS)} innovatively adopts RL to dynamically explore the interactions among different supervised layers, thereby optimizing the distribution of the balancing weight of supervision loss under a task-specific objective. This novel auto-weighting mechanism guides the deep supervision module to scrutinize the contributions of different supervised levels to favor highly semantically discriminative myocardial pathology features.

\textbf{Pixel-wise Attention Module (PAM)} learns an attention-aware matrix to emphasize the meaningful region of scar and edema. PAM explores the corresponding feature maps of down-sampling and up-sampling paths to integrate their spatial relationship. Therefore, it perceives the location and size of the desirable pathology from low-level to high-level comprehensively.

\subsection{Cascaded Anatomical Segmentation Network (CASN)}
\label{CASN}
\begin{figure}[!t]
	\centering
	\includegraphics[scale=.5]{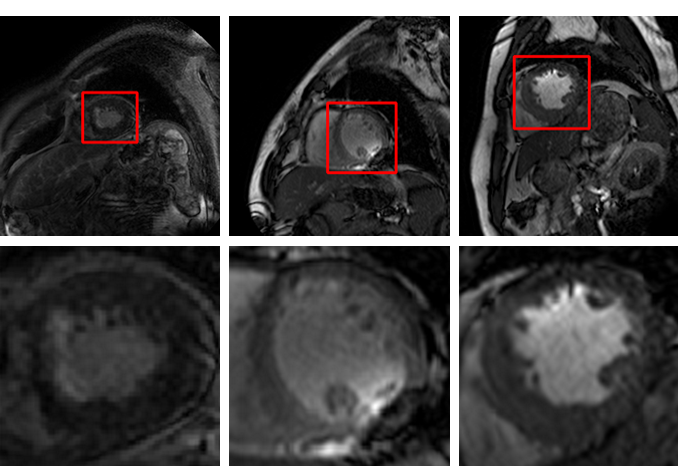}
	\caption{An example of the clipped rectangular region in the multi-sequence CMR. The left ventricle area is framed by a red rectangle, and the cropped image is shown in the second row.}
	\label{shujuyuchuli}
\end{figure}

We first design a Cascaded Anatomical Segmentation Network, called CASN, to automatically delineate LV epicardial contours containing the blood pool and the myocardium on which the scar and edema are distributed. As illustrated in
Fig.
\ref{zongkuangtu}
, our proposed CASN consists of two vanilla U-Net in a cascaded mode and takes the bSSFP, LGE and T2-weighted CMR sequences as input in an early-fusion manner. Given multi-sequence CMR images
$I_1 = (I_{bSSFP}, I_{LGE}, I_{T2})$  the first vanilla U-Net is designated to learn a mapping to indicate the region containing the LV coarsely. Based on the coarse segmentation, we can crop the rectangular region of interest (ROI) $I_{1t} = ( I_{bSSFP,t}, I_{LGE,t}, I_{T2,t} )$ from the images accordingly as the input to reduce the false positive of LV cavity segmentation for the followed U-Net. Fig.
\ref{shujuyuchuli} presents an example of the clipped rectangular region in the multi-sequence CMR. Similarly, the second vanilla U-Net adopts the same architecture to segment the LV anatomical structure more accurately, contributing to data balance for the followed scar and edema segmentation. We directly concatenate the ROI from the bSSFP, LGE and T2-weighted CMR to form 3-channel input. The CASN is trained under the Dice loss since Dice score coefficients (DSC) can mitigate the data imbalance in segmentation
\citep{yang2018towards}. The DSC for the LV segmentation of channel $c$ is expressed as below:
\noindent
\begin{equation}
	DSC_{c}=\displaystyle\frac{ 2\sum\limits_{i=1}^Np_{ic}g_{ic} + \epsilon  }{  \sum\limits_{i=1}^N(p_{ic}+g_{ic})+ \epsilon  }
\end{equation}
where $p_{c} \in \left\{ 0,1 \right\} $ and $g_{c} \in  \left\{ 0,1 \right\}$ represent the ground truth label and the predicted label, respectively. The N is the total number of pixels in the cropped region. The $\epsilon$ term is used here to ensure the loss function stability by avoiding the numerical issue of dividing by 0. This CASN network finally learning a mapping $f_{\theta}$ from $I_{1}$ to a binary mask of the LV region. The candidate anatomical structure can be extracted from the multi-sequence CMR as below: 
\begin{equation}
	A_{1t} = I_{1t} \otimes f_{\theta}(I_{1})
\end{equation}
where $\otimes$ is element-wise multiplication. 

The small regions of scar and edema mean that background pixels absolutely dominate foreground pixels. However, most of the learning-based methods bias predictions towards the majority class because of imbalanced data distribution, especially in segmenting small structures
 \citep{koziarski2018convolutional}. The CASN can explore candidate anatomical structure only consisting of the blood pool, myocardium, scar, and edema. Therefore, the foreground pixels ratio (scar and edema) can become larger, mitigating the data imbalance in segmenting the scar and edema. The output from CASN can also provide the prior spatial information to boost the following myocardial pathology segmentation performance. On the other hand, the bSSFP, LGE, and T2-weighted CMR emphasize different salient anatomy structures with noticeable low-level features, such as intensity distribution, respectively. The early fusion in CASN makes it possible to exploit the correlation and interactions between low-level features among various CMR sequences to enhance performance.

\subsection{Auto-weighted Supervision Attention Network (ASAN)}
\label{ASAN}
The bSSFP cine sequence captures cardiac motions and presents clear boundaries but almost no myocardial pathology information. Therefore, we only concatenate the LGE and T2-weighted CMR masked using LV segmentation to extract the pertinent myocardial pathology information with our developed auto-weighted supervision attention network, called ASAN. As shown in Fig.
\ref{zongkuangtu} , the ASAN is actually a variant of U-Net with pixel-wise attention and deep supervision submodules, in which the encoder pathway (down-sampling path) is similar to the vanilla U-Net to extract high-level semantic features layer by layer. On the other hand, we follow the deep supervision mechanism to make predictions from different layers in the decoder pathway (i.e., up-sampling path). We also employ a newly RL-based auto-weighted mechanism to dynamically learn the optimal weights among various supervised layers. Moreover, the pixel-wise attention submodules are embedded into the skip connection between the encoder and decoder at each layer for concentrating the meaningful small pathological structure.

Edema is due to water accumulation in the injured myocardium and reaches maximal over the first week after MI. Meanwhile, myocardial scarring is fibrosis resulting from irreversible trauma to the cardiac tissue over the course of several weeks. Moreover, the edema is usually larger than myocardial scarring in MI. Even though both LGE CMR and T2-weighted CMR can reveal myocardial injury with regional hyperenhancement, it is still difficult to distinguish scar from edema because of the ambiguous boundary between these two dynamic changing injured tissues
\citep{das2019role}. Therefore, instead of recognizing scar and edema simultaneously using one unified network, we separately train two independent ASAN network with early fusion: one network identifies three categories, namely background, scar, and ring-shaped myocardium from multi-sequence CMR images, while another ASAN aims to extract myocardial pathology (scar and edema) from ring-shaped myocardium and background. With the input of the concatenation of LGE and T2-weighted CMR, these two networks can emphasize salient features on different pathological structures from LGE and T2-weighted CMR, better for the data imbalance and ambiguity between scars and edema. Besides, the ring-shaped myocardium can afford additional spatial prior information to refine these pathology regions. The output from these ASANs is merged to obtain the pathology region containing both scar and edema. Finally, one ensemble was generated to improve the accuracy of the final predictions further.

\subsubsection{Deep Auto-weighted Supervision(DAS)}
\label{Deep Supervision Attention U-Net}
When given limited training data in the small organ segmentation, the notorious appearance of gradients vanishing or exploding could hinder the convergence of the training process. Deep supervision can effectively cope with this optimization problem since it establishes shortcut connections from the loss to the weights in hidden layers, improving the prorogation of gradient flows within the network 
\citep{lee2015deeply}. It can also directly guide the intermediate feature-maps to favor highly semantically discriminative features at each image scale, further boosting its generalization capability  
\citep{dou20173d}. Therefore, we employ deep-supervision to ensure that PAM can influence the responses to learn more powerful and representative features at different scales within the network and simultaneously speed up the optimization process. 

\textbf{Loss Function}: Although the Dice loss function can partly mitigate the data imbalance problem, its gradient disappearance problem can slow the convergence efficiency. The Dice loss can also not explicitly control the trade-off between false positive (FP) and false negative (FN). In contrast, the cross-entropy loss can force the model to learn poorly classified pixels better and to train smoothly, alleviating the problems of slow-down learning and the vanishing gradient. Therefore, we combine the Dice and the cross-entropy loss (DCL) to achieve a better segmentation in the deep supervision attention U-Net. We then formulate the hybrid loss at layer $l$ as below :
\begin{equation}
	L^{l} = L_{Dice}^{l} + \lambda_{CE}L_{CE}^{l}
\end{equation}
Where $L_{Dice}^{l}$ and $L_{CE}^{l}$ represent the Dice loss and cross-entropy loss, and $\lambda_{CE}$ is the coefficient of cross-entropy loss. Finally, we train the ASAN using the backpropagation algorithm by minimizing the following loss function:
\begin{equation}
	L =\sum\limits_{l \in M}\alpha_{l}L^{l}
\end{equation}
where $M$ is the set of the index of all supervision layers, and $\alpha_{l}$ is the balancing weights of each layer supervision loss, which satisfies $\sum\limits_{l \in M}\alpha_{l} = 1$.

\begin{figure*}[!t]
	\centering
	\includegraphics[width=\textwidth]{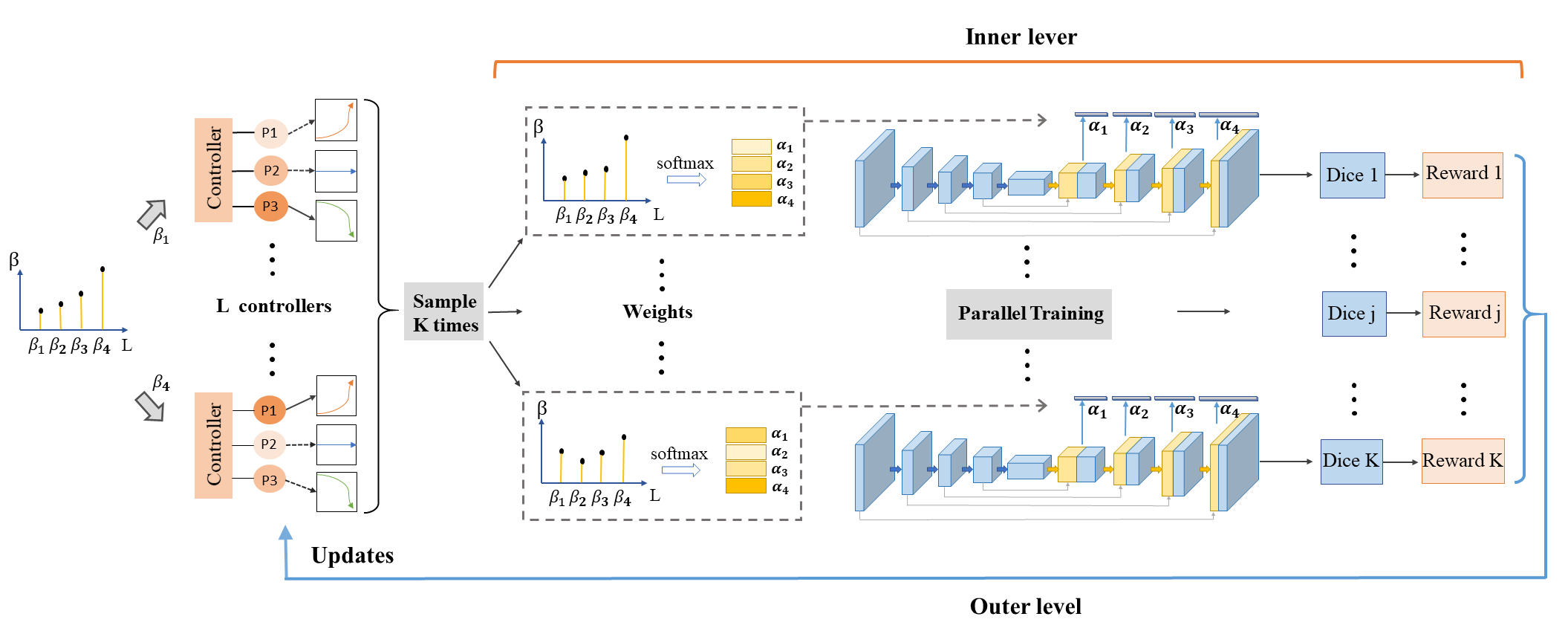}
	\caption{The flow diagram of the proposed deep auto-weighted supervision (DAS) module.}
	\label{DAS}
\end{figure*}

\textbf{DAS}:
The performance of deep supervision is dependent on the distribution of the balancing weight of supervision loss, which is an inherited inconsistency and uncertainty among different tasks. Like multi-modality learning, ideal deep supervision should make predictions by emphasizes one or two supervision layers from all the available information. However, their distribution often follows a fixed probability distribution, and those weights of supervision loss at hidden layers decay during deep supervision training. It is impossible to define these hyperparameters heuristically as we can hardly identify the more essential layers for different tasks in advance. Based on our preliminary work in the multi-modality automating meta-parameters
\citep{wang2020auto}, we advance the dynamical optimization of the balancing weight of supervision loss using an RL framework, where an agent, in its current state $S$, interacts with the environments $E$ by making successive actions $a\in A$ that maximizes the expectation of reward to tune these model hyperparameters.

In this study, we take a group of multi-modality CMR inputs as the environment and define states $S$
as the set of weights corresponding to various supervision layers. These weights, represented by $\left\{\alpha_{k}\right\}$, are predicted by applying a softmax function to four real-valued $\left\{{\beta_{k}}\right\}$: $\alpha_{k} = \frac{exp(\beta_{k})}{\sum_{k=1}^{4}{exp(\beta_{k})}}$. The system continuously optimizes the weighting among different deep supervision by employing the agent to interact with the environment to maximize the reward. Each agent called the controller then seeks the appropriate action by employing a softmax function on controller parameters $\left\{\theta_{k}\right\}$. To be specific, we define the action space in an additive way, in which
$\left\{\Delta\beta_{k}\right\}$ has three options $\left\{-0.15,0,0.15\right\}$, corresponding to different output changes determined by the controller parameters $\left\{\theta_{k}\right\}$. We use three random values from the Gaussian distribution with a variance of 0.001 to initialize the controller parameters, making the initial difference among the three parameters minimal. Each controller then takes the segmentation accuracy calculated in the latest few epochs as the scalar reward $R$ to explore whether the agent is moving towards the preferred target for weights optimization.

This optimization task is a standard bilevel optimization problem. We aim to both maximize the reward w.r.t. the controller parameters $\left\{{\theta}\right\}$, as well as minimize the objective loss of the network w.r.t. model parameters $\left\{{\omega}\right\}$. We alternately train this bi-level model to update $\left\{{\theta}\right\}$ and $\left\{{\omega}\right\}$.  As illustrated in Fig.
\ref{DAS}, at the inner level, we sample K times from the parameter space to generate K identical networks with different deep supervision balancing weights $\left\{\alpha_{1:4}^{t,1:K}\right\}$. These $K$ networks are independently trained on the training set for one epoch to update their corresponding network parameters, respectively. In contrast, we seek the accuracy of all verification models to update $\left\{{\theta_{i}}\right\}$ for each controller at the outer level of optimization. Each controller corresponding to one supervision layer is independent of each other, which are updated iteratively using the REINFORCE rule 
\citep{williams1992simple} as below :

\begin{equation}
	\theta_{i}^{t+1}=\theta_{i}^{t} +\eta \frac{1}{K}\sum\limits_{j=1}^KR_{j}\cdot\triangledown_{\theta}log(g(\beta^{t,j}))
\end{equation}
\begin{equation}
	\theta_{i}^{t+1}=\theta_{i}^{t} +\eta \frac{1}{K}\sum\limits_{j=1}^KR_{j}\cdot\triangledown_{\theta} \sum\limits_{i=1}^4log(p(\beta_{i}^{t,j})),
	\label{equation8}
\end{equation}
where $\eta$ is the learning rate, $g(\beta^{t,j})$ represents the joint probability distribution of different deep supervision loss weights in the  $j$-$th$ sample at $t$-$th$ epoch. We assume that all the weights are independently distributed, based on which we can get  $g(\beta^{t,j}) = \prod(p(\beta_{i}^{t,j}))$.
 The formulation of the reward is then defined as:
\begin{equation}
	R_{j}=(\epsilon_{j}+0.04)^3 
	\label{equation9}
\end{equation}
where $\epsilon_{j} \in [0,1]$ is the Dice metric calculated on the validation set with $j-th$ network. The cubic function in equation
\ref{equation9} is utilized to strengthen the reward signal. Moreover, we add a baseline term $B^{t}$
in equation 
\ref{equation8} to reduce the variance, improving the stability and convergence of the model. We rewrite the update of the controller as follows:
\begin{equation}
    \theta_{i}^{t+1}=\theta_{i}^{t}+\eta_{i}\frac{1}{K}\sum\limits_{j=1}^K(R_j-B^t)\cdot\triangledown_{\theta} \sum\limits_{i=1}^4log(p(\beta_{i}^{t,j}))
	\label{equationbt}
\end{equation}
\begin{equation}
	B^{t}=(1-\gamma)B^{t-1}+\gamma\cdot(\frac{1}{K}\sum\limits_{j=1}^KR_{j})
\end{equation}
where $\gamma$ is a constant. Specifically, we add a baseline term $B^{t}$ in equation \ref{equationbt} to restrict the reward variance, improving the stability and convergence of the model. According to \citep{williams1992simple}, using the baselines in the enhanced comparison strategies can greatly increase convergence speed. The simulation results in \citep{dayan1991reinforcement} also supported that the usage of a baseline value can positively affect the convergence speed of reinforcement learning. It is worth noting that we employ $K$ networks independently to sustain necessary exploration for other actions during training, acquiring the well-learned policy. On the other hand, we exploit the model and actions with the highest validation performance, thus alleviating the general exploitation-exploration dilemma in the RL-based framework.

\subsubsection{Pixel-wise Attention Module (PAM)}
\label{PAM}
The CASN mitigates the data imbalance using the prior anatomy information. Nonetheless, it remains challenging to reduce false-positive predictions for small scar and edema that show large shape variability across patients. These hard pixels from scar and edema are also often ambiguous, bringing additional difficulty in segmentation. Attention mechanism has been recently increasingly used to make the model focus on the most latent region \citep{roy2018recalibrating,mei2021automatic}. Inspired by the Attention Gate (AG) to recalibrate the learned feature maps
\citep{oktay2018attention} , we plug a pixel-wise attention module (PAM) into the skip connections from the encoder to the decoder path. PAM generates attention-aware features using the element-wise multiplication of input feature-maps and attention mask, which learns to focus on the meaningful small pathological structure by identifying salient image regions and inhibiting irrelevant responses. Therefore, PAM can preserve the activations according to the specific anatomy progressively.

Fig.
\ref{zongkuangtu} presents the schematic diagram of the PAM module. The PAM employs a similar gating vector between the encoder to the decoder path to derive the attention mask. This gating vector aggregates information from multiple imaging scales to prune feature responses in irrelevant background regions by measuring the dependency between the corresponding feature maps of down-sampling and up-sampling paths. Instead of down-sampling the encoder path in the original gating vector, we only resample the decoder path at each layer of the proposed 2D U-Net to avoid possible details loss, especially for the tiny scar and edema regions. The PAM derives the attention mask from suppressing feature responses in irrelevant regions by measuring the dependency between the corresponding feature maps of down-sampling and up-sampling paths. The PAM first uses element-wise addition to merge feature maps to obtain the intermediate features. A projection operation is then employed to match the channel dimensions using the BN-RELU-1×1-convolution operation. Finally, the PAM utilizes a sigmoid function to obtain the attention mask as a probability distribution to indicate focus regions. We adopt additive attention in PAM since it can achieve higher accuracy than multiplicative attention
\citep{shen2017disan}. We can compute scalar attention value $\alpha_{c}^{l}$  at each pixel corresponding to the channel $c$ feature-maps in layer $l$ using the below additive attention formulation:
\begin{equation}
	\alpha_{c}^{l} = \delta(W^{T}(\psi(W_{c,l,d}^{T}d_{c,l}+W_{c,l,u}^{T}u_{c,l}+b_{1}))+b_{2})
\end{equation}
\begin{equation}
	\widehat{u}_{c,l} = \alpha_{c}^{l} \cdot u_{c,l} 
\end{equation}
where $\psi$ corresponds to a ReLU function, $\delta$ is the sigmoid activation function, $b_{1}$, $b_{2}$ are the bias of convolution, and  $d_{c,l}$ and $u_{c,l}$ are the feature maps from both down-sampling and up-sampling paths, respectively. 

The plug-in PAM model automatically learns to highlight the salient features without additional supervision (only the related activations are merged with the concatenation operation), see Fig.
\ref{zongkuangtu}. We can eliminate irrelevant and noisy responses in skip connections by using the coarse-scale information when plugging PAM instead of direct concatenation. Moreover, the PAM embedded in each skip connection better aggregates a set of information from the previous scales, increasing the grid-resolution and achieving better-focusing performance. Furthermore, the PAM serves as a gradient update filter since gradients from the irrelevant area are restricted during backpropagation. Then PAM can allow model parameters to update mainly based on the region according to the specific tasks during training.

\subsubsection{Ensemble learning}
It is generally difficult to know a priori the best prediction solution given the limited training data in this study. The ensemble learning methods have gained popularity due to their superior prediction performance in practice \citep{ju2018relative}. To exploit the dynamic learning advantage sufficiently, we further employ ensemble learning to enhance the segmentation results. Since bagging, the most popular homogeneous ensemble learning technique, has shown to reduce uncertainty in models and increase generalization capabilities \citep{zhou2009ensemble}, we adopt bagging to ensemble the combination of the base learners. In this study, we slightly modify initial weights corresponding to learning-based supervised layers to train multiple homogeneous base-learner equipped with the auto-weighting. During inference, the corresponding pixel-wise prediction is acquired for each base-learner. Finally, we derive the segmentation result pixel-by-pixel by majority voting.

\section{Experiments}
\label{sec4}
\subsection{Datasets}
We assessed the proposed method on a publicly available dataset from Myocardial pathology segmentation combining multi-sequence CMR (MyoPS 2020), supported by the Medical Image Computing and Computer-Assisted Intervention (MICCAI) in 2020. The MyoPS 2020 dataset consists of multi-sequence CMR, including bSSFP, LGE, T2, from 45 patients. The LGE CMR was a T1-weighted, inversion-recovery, gradient-echo sequence to cover the main body of the ventricles, the T2 CMR was a T2-weighted, black blood spectral presaturation attenuated inversion-recovery (SPAIR) sequence generally consisting of a small number of slices, and the bSSFP CMR was a balanced steady-state, free precession cine sequence covering the full ventricles from the apex to the basal plane of the mitral valve. Each CMR sequence generally comprises 2 to 6 slices with in-plane resolution $0.75\times0.75$ mm and in-plane size ranged from $412\times408$ to $512\times515$. The three multi-slice sequences under the same cardiac phase were all breath-hold acquired in the ventricular short-axis views and were aligned into a common space using the MvMM
\citep{zhuang2019multivariate, zhuang2016multivariate}. The slice-by-slice manual delineations from three independent and well-trained observers were then averaged to obtain the standard golden segmentation as the ground truth. The details of the three CMR sequences dataset refer to 
\citep{zhuang2019multivariate}.

\subsection{Implementation details}
We ran all experiments on a machine with an NVIDIA TITAN X(PASCAL) GPU to evaluate our proposed model with the U-Net
\citep{ronneberger2015u}, U-Net++
\citep{zhou2018unet++}, and nnU-Net
\citep{isensee2018nnu}. We employed the same coarse-to-fine strategy to implement all these methods in our codes with Pytorch to ensure a consistent comparison with our proposed model. All original images were first cropped to $384\times 384$ based on the image center as the input to our proposed CASN network to identify the LV region. The ROI was then cropped using the LV epicardial contour with around ten-pixels outside margins and resized to a dimensional of $128\times 128$ to input the second fine segmentation network ASAN, U-Net, U-Net++, and nnU-Net, to achieve myocardial pathology segmentation. All these fine segmentation networks adopted the same hybrid loss, the weighted sum of dice loss and cross-entropy loss with $\lambda_{CE} = 0.25$. Moreover, we evaluated the proposed method with the results in state-of-art literature about the public MyoPS 2020 dataset 
\citep{myops2020}.

We followed MyoPS 2020 dataset guideline to use 25 cases as the training set and the remaining cases as the testing set. We further divided the training set into 23 cases for training and 2 cases for validation, respectively. We trained the controllers initialized with Gaussian distribution ($\mu = 0$, $\sigma = 0.001$) and applied different augmentation strategies, including flipping, rotation (up to $10^{\circ}$), and brightness and saturation adjustment for all models. All networks are trained for 200 epochs using Adam optimizer with a learning rate of 0.001 and a mini-batch size of 16. In particular, after training the ASAN network 150 epochs with fixed-weight deep supervision, we begin to employ the auto-weighted deep supervision framework to search for the optimum weights, improving training stability. We began the auto-weighted stage by setting $\beta_{k}$ as $[0.5,0.5,0.5,1]$ to initialize the weight of each supervision layer and dynamically optimized these weights accordingly until 50 epochs in total. During training, we set the sample number $K = 10$ and the trade-off parameter $\gamma = 0.99$ and update the controller parameters every epoch using an Adam optimizer with a learning rate of 0.001. We followed previous automating meta-parameters optimization methods \citep{baker2016designing,li2019lfs,zoph2016neural} to optimize the network parameters and the corresponding network hyperparameters (i.e., the weighting of different supervision layers) with the training dataset and validation dataset, respectively. It was noted that the training time for each epoch with and without reinforcement learning is 16.8 and 1.54 seconds, respectively. This ten-times time difference is attributed to training ten sample networks with different deep supervision coefficient networks to explore the optimal weights under the RL-based framework.

\subsection{Evaluation Criteria}
We evaluate our proposed method using four criteria, including Dice similarity coefficient, Jaccard coefficient(JC), Hausdorff Distance(HD) of Boundaries, and average symmetric surface distance (ASD), from both the region and boundary similarities. The often-used metric in validating region similarities is the Dice similarity coefficient ($Dice =2\frac{A \cap B} {A + B} $), measuring a spatial overlap index between segmentation and ground truth. Another region similarities index ($Jaccard =\frac{A \cap B} {A \cup B} $) emphasizes coordinates similarity between finite sample sets. On the other hand, the HD indicates the worst labeling cases by using the greatest of all the distances from a point in one set to the closest point in the other set but being sensitive to boundary outliers
\citep{kamnitsas2017efficient}. The ASD metric describes the average distance from segmentation to ground truth to evaluate shape fidelity, which is stable and less sensitive to outliers than the HD.

\section{Results and Discussion}
\label{sec5}
\subsection{Ablation Experiments}
We run a number of ablation experiments to analyze the influence of each submodule in our proposed method. We first build the baseline model using the vanilla U-Net with four-layer architecture and gradually incorporate each submodule discussed in Section
\ref{sec3} into the baseline model for further ablation comparison. Moreover, we investigated the effect of the CMR data by input different combinations of CMR sequences to the ASAN network. We further scrutinized the necessity of two independent ASAN networks by applying one single ASAN network to identify background, scar, myocardial pathology (scar and edema), and ring-shaped myocardium simultaneously. All these experiments are conducted using the same aforementioned training configurations.

\begin{figure*}[!t]
	\centering
	\includegraphics[width=\textwidth]{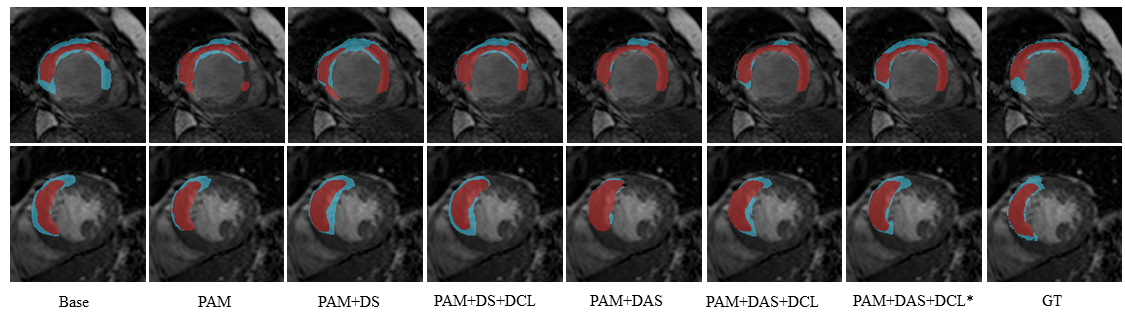}
	\caption{Visualize the segmentation results in the MyoPS dataset by using different training combinations. Ensemble learning is denoted by superscript *. Among them, scar and edema are shown in red and blue, respectively.}
	\label{ablation}
\end{figure*}

 Table 
\ref{tab1}
 presents the average Dice, JC, HD, and ASD from ablation experiments. It quantitatively demonstrates that each submodule contributes to performance improvement, verifying the superiority of our proposed method. Compared to the baseline model, using the attention and deep supervision module alone can achieve a higher Dice and Jaccard index with about a maximum 0.5\% improvement, but failing to enhance the boundary similarity. We observe higher performance in both region and boundary similarities when employing the hybrid loss of Dice loss and the cross-entropy loss. Note that using the hybrid loss achieves a noticeable performance improvement in boundary similarity, e.g., it reduces about 6.5mm in HD and 1.3mm in ASD in segmenting scar. The small scar and edema regions imply data imbalance, preferring the Dice loss to derive the segmentation map
 \citep{yang2018towards}. However, the Dice loss fails to explicitly penalize the deviation at each pixel, degrading the performance in the boundary similarity. Table 
 \ref{tab1}
validates that using the cross-entropy function can partly address this limitation. Moreover, Fig.
\ref{ablation}  visualizes an example for illustrating the segmentation results of scar and edema from the mentioned ablation study. The comparison with and without DCL also suggests that the results of hybrid loss are closer to the ground truth than only using the Dice loss. These results suggest that learning hard classified pixels through the hybrid loss is important in myocardial pathology segmentation.

\begin{table*}[!t]
	\small
	\setlength{\belowcaptionskip}{4pt}
	\caption{\label{tab1}Summary of the quantitative evaluation results of scar and edema segmentation in the MyoPS dataset. 
		Ensemble learning is denoted by superscript $^{*}$.}
	\centering
	\renewcommand\arraystretch{1.5}
	\resizebox{\textwidth}{!}{
		\begin{tabular}{c|cccc|cccc}
			\hline
			\multirow{2}{*}{Method}&\multicolumn{4}{|c|}{Scar}&\multicolumn{4}{|c}{Scar + Edema}\\
			\cline{2-9}
			&Dice $\uparrow$&JC $\uparrow$&HD (mm) $\downarrow$&ASD (mm) $\downarrow$&Dice $\uparrow$&JC $\uparrow$&HD (mm) $\downarrow$&ASD (mm) $\downarrow$\\
			\hline
			Single & 0.636±0.256& 0.508±0.230&19.24±13.52&2.32±2.86&0.704±0.100&0.555±0.132&21.02±11.22&2.12±2.56\\
			Base & 0.628±0.251 & 0.500±0.232 &20.20±15.31 & 1.92±3.43 & 0.685±0.108 & 0.531±0.123 & 23.41±10.92 & 2.11±1.61 \\
			PAM & 0.631±0.249 & 0.503±0.233 & 22.48±17.17 & 2.44±3.11 & 0.693±0.113 & 0.541±0.128 & 23.32±11.14 & 2.16±2.00 \\
			PAM+DS & 0.634±0.251 & 0.506±0.234 & 21.50±14.62
			& 2.84±4.70 & 0.708±0.136 & 0.563±0.147& 19.54±10.74 & 2.06±3.62\\
			PAM+DS+DCL & 0.637±0.261 & 0.514±0.245 & 14.95±12.41 & 1.70±2.40& 0.714±0.129& 0.570±0.151 & 14.71±9.71& 1.44±2.69 \\
			PAM+DAS  & 0.654±0.239 & 0.524±0.222 & 17.02±14.06 & 1.45±2.00& 0.711±0.119 & 0.564±0.136 & 17.33±10.95& 1.46±1.89 \\
			PAM+DAS+DCL  & 0.658±0.253 & 0.535±0.240 & 14.13±11.90 & 1.52±2.28 & 0.720±0.126	& 0.577±0.148 &\textbf{14.12±9.24} & 1.37±2.15\\
			PAM+DAS+DCL$^{*}$ &\textbf{ 0.678±0.242} & \textbf{0.555±0.228} & \textbf{10.23±9.53} &\textbf{ 0.82±1.48 }& \textbf{0.735±0.111} &\textbf{0.592±0.132} & 15.34±9.87 & \textbf{1.18±1.66} \\
			\hline
	\end{tabular}}
\end{table*}

\begin{table*}[!t]
	\small
	\setlength{\belowcaptionskip}{4pt}
	\setlength{\tabcolsep}{6mm}
	\caption{\label{tabpvalue1}Paired T-test (p-value) between baseline and different modules in the MyoPS dataset.}
	\centering
	\renewcommand\arraystretch{1.5}
	\begin{tabular}{c|cc|cc}
		\hline
		\multirow{2}{*}{}&\multicolumn{2}{|c|}{Scar}&\multicolumn{2}{|c}{Scar + Edema}\\
		\cline{2-5}
		&Dice &HD &Dice &HD\\
		\hline
		PAM vs. Baseline & p=0.42 & p=0.19 & p=0.22 & p=0.47 \\
		DAS vs. Baseline & p\textless0.05 & p\textless0.05 & p\textless0.05 & p\textless0.05 \\
		PAM+DAS vs. Baseline & p\textless0.05 & p\textless0.05 & p\textless0.05
		& p\textless0.001 \\
		Ensemble vs. PAM+DAS & p\textless0.05 & p\textless0.05 & p\textless0.001&p\textless0.001 \\
		\hline
	\end{tabular}
\end{table*}

\begin{table}[!t]
	\small
	\setlength{\belowcaptionskip}{4pt}%增大表格与标题的距离
	\setlength{\tabcolsep}{7mm}
	\caption{\label{tabmultimodel}The results of the myocardial pathology segmentation with different CMR sequence  in the MyoPS dataset.}
	\centering
	\renewcommand\arraystretch{1.5}
	\begin{tabular}{c|c|c}
		\hline
		\multirow{2}{*}{}&Scar&Scar + Edema\\
		\cline{2-3}
		&Dice $\uparrow$&Dice $\uparrow$\\
		\hline
		bSSFP & 0.406±0.232 & 0.492±0.140\\
		LGE & 0.615±0.266 & 0.647±0.124\\
		T2 & 0.455±0.247 & 0.636±0.141\\
		LGE+T2 & 0.658±0.253 & 0.720±0.126\\
		\hline
	\end{tabular}	
\end{table}

Meanwhile, Table
\ref{tab1} illustrates that auto-weighted supervision can also boost segmenting scar and edema. For instance, the scar segmentation results using auto-weighted supervision increase by about 2.0\% in the Dice and JC, and decrease about 20\% in HD and ASD compared with fixed-weighted results. As shown in Fig.\ref{ablation}, the segmentation visualization results also look reasonable, and the boundary is more consistent with the ground-truth. Furthermore, it quantitatively proves that combining all these submodules can further promote scar and edema segmentation performance. Overall, the ASAN model achieves a mean of 65.8\% and 72.0\% in Dice for scar and edema, respectively, which means it has a good ability to separate each myocardial pathology pixel correctly from the background. On the other hand, a single ASAN model identifying scar and scar\&edema together can have much less improvement than using a dual ASAN model to identify scar and scar\&edema, respectively (Table \ref{tab1}, line1). Finally, our proposed model, through ensemble, performs most superiority in segmenting myocardial pathology. It shows that the final model obtained the highest Dice among all ablation tested models. Fig.
\ref{ablation} also illustrates that it completes a good segmentation despite the complicated appearance of scar and edema. Moreover, we have used the paired t-test to evaluate whether the performance improvement is statistically significant (Table \ref{tabpvalue1}). These results indicated that the improvement in Dice and HD was significant (p\textless0.05) when employing the DAS module, validating its necessity in our proposed AWSnet. By contrast, both Table \ref{tab1} and Table \ref{tabpvalue1} suggest that the PAM module has less contribution to the performance improvement. Besides, ensemble learning can commit a statistically significant improvement in scar and edema segmentation. 
	
Furthermore, the ablation study in Table \ref{tabmultimodel} shows the myocardial pathology segmentation with different CMR sequence inputs. It is clear that the results using the single bSSFP sequence results are much worse than the others, proving its limitation on presenting myocardial pathology information. By contrast, the model using LGE CMR achieved much higher segmentation accuracy in both scar and edema. Also, as the T2-weighted CMR mainly depicts the edema region, it was more beneficial for edema segmentation. These results are consistent with the enhancement patterns of these CMR sequences reported in the literature, implying that combining T2-weighted and LGE CMR might boost segmenting both scar and edema \citep{flett2011evaluation,kim2009cardiovascular}.

\begin{table*}[!t]
	\small
	\setlength{\belowcaptionskip}{4pt}
	\caption{\label{tab2}Summary of the quantitative evaluation results of scar and edema segmentation in the MyoPS dataset.}
	\centering
	\renewcommand\arraystretch{1.5}
	\resizebox{\textwidth}{!}{
		\begin{tabular}{c|cccc|cccc}
			\hline
			\multirow{2}{*}{Method}&\multicolumn{4}{|c|}{Scar}&\multicolumn{4}{|c}{Scar + Edema}\\
			\cline{2-9}
			&Dice $\uparrow$&JC $\uparrow$&HD (mm) $\downarrow$&ASD (mm) $\downarrow$&Dice $\uparrow$&JC $\uparrow$&HD  (mm) $\downarrow$&ASD (mm) $\downarrow$\\
			\hline
		   	U-Net & 0.628±0.251 & 0.500±0.232 &20.20±15.31 & 1.92±3.43 & 0.685±0.108 & 0.531±0.123 & 23.41±10.92 & 2.11±1.61 \\
			U-Net++ & 0.646±0.247 & 0.520±0.234 & 18.03±15.85 & 2.23±3.33
			    & 0.684±0.108 & 0.530±0.124 & 23.38±11.68 &2.54±2.17 \\
			nnU-Net & 0.634±0.251 & 0.496±0.235 & 20.90±15.52 & 2.52±4.16 & 0.674±0.136 & 0.523±0.146 &22.20±10.62 & 2.14±2.70\\
			Ours  &\textbf{0.658±0.253} & \textbf{0.535±0.240} & \textbf{14.13±11.90} &\textbf{1.52±2.28}& \textbf{0.720±0.126} &\textbf{0.577±0.148} & \textbf{14.12±9.24} & \textbf{1.37±2.15} \\
			\hline
	\end{tabular}}
\end{table*}

\begin{table}[!t]
	\small
	\setlength{\belowcaptionskip}{4pt}
	\setlength{\tabcolsep}{3.6mm}
	\caption{\label{tabpvalue2}Paired T-test (p-value) between AWSnet and other networks in the MyoPS dataset.}
	\centering
	\renewcommand\arraystretch{1.5}
	\begin{tabular}{c|cc|cc}
		\hline
		\multirow{2}{*}{}&\multicolumn{2}{|c|}{Scar}&\multicolumn{2}{|c}{Scar + Edema}\\
		\cline{2-5}
		&Dice &HD &Dice &HD\\
		\hline
		U-Net & p\textless0.05 & p\textless0.05 & p\textless0.001 & p\textless0.001 \\
		U-Net++ & p=0.11 & p\textless0.05 & p\textless0.001 & p\textless0.05 \\
		nnU-Net & p\textless0.05 & p\textless0.001 & p\textless0.001
		& p\textless0.05 \\
		\hline
	\end{tabular}
\end{table}

\begin{figure*}[!t]
	\centering
	\includegraphics[scale=0.65]{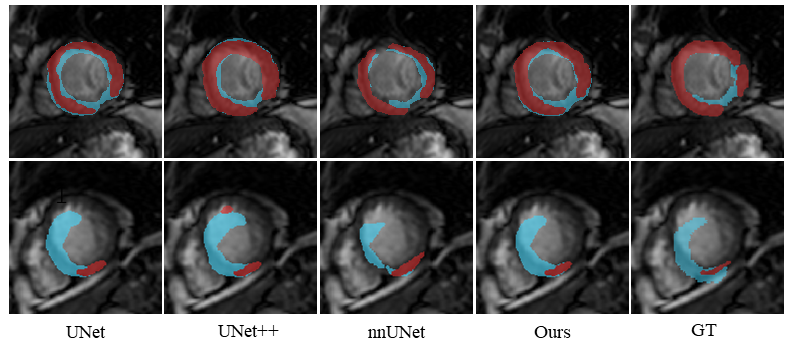}
	\caption{The visualization of segmentation results in the MyoPS dataset compared with other classic methods.}
	\label{fig6}
\end{figure*}

\begin{table*}[!t]
	\small
	\setlength{\belowcaptionskip}{4pt}
	\setlength{\tabcolsep}{6mm}
	\caption{\label{tab3}Results of the MyoPS 2020 challenge of different methods (Dice). Only the top-10 methods are included in this table. For full rankings, please kindly refer to the Challenge website \href{http://www.sdspeople.fudan.edu.cn/zhuangxiahai/0/myops20/}{http://www.sdspeople.fudan.edu.cn/zhuangxiahai/0/myops20/} .}
	\centering
	\renewcommand\arraystretch{1.5}
	\begin{tabular}{c|ccc}
		\hline
		TOP10 &Scar&Scar + Edema&Average\\
		\hline
        Team1 \citep{zhai2020myocardial} & 0.672±0.244 & 0.731±0.107 & 0.702\\
        Team2 \citep{martin2020stacked} & 0.666±0.242 & 0.698±0.129 & 0.682\\
        Team3 \citep{zhang2020efficientseg} & 0.647±0.279 & 0.709±0.122 & 0.678\\
        Team4 \citep{li2020cms} & 0.581±0.268 & 0.725±0.110 & 0.653\\
        Team5 \citep{liu2020two} & 0.635±0.290 & 0.669±0.114 & 0.652\\
        Team6 \citep{zhang2020multi} & 0.613±0.238 & 0.674±0.131 & 0.644\\
        Team7 \citep{ankenbrand2020exploring}& 0.620±0.240 & 0.665±0.137 & 0.643\\
        Team8 \citep{li2020dual}& 0.605±0.263 & 0.656±0.138 & 0.631\\
		Team9 \citep{arega2020automatic}& 0.565±0.272 & 0.664±0.150 & 0.615\\
		Team10 \citep{zhao2020stacked}& 0.586±0.266 & 0.639±0.141 & 0.613\\
		\hline
		Ours & \textbf{0.678±0.242} & \textbf{0.735±0.111} & \textbf{0.707}\\
		\hline
	\end{tabular}
\end{table*}

\subsection{Comparison Experiments}
As there is little study focusing on myocardial pathology segmentation using multi-sequence CMR, we first implemented the vanilla U-Net method and its state-of-art two variants for the performance comparison. We chose the U-Net++ and nnU-Net, strengthening the segmentation results by introducing nested and dense skip connection and a robust and self-adapting framework based on 2D and 3D U-Nets. We used the same hyper-parameters in this experiment for consistency. The results are reported in Table 
\ref{tab2}. We also provide some visualization results for more in-depth discussion. Moreover, we further compared the proposed method with the state-of-art literature about the public MyoPS 2020 dataset. Note that only the Dice metric is available in the results from the public MyoPS 2020 challenge.

Table
\ref{tab2} tabulates the quantitative comparison results for both the scar and edema segmentation. Our method achieves 67.8\% and 73.5\% in Dice, implying the ability to cope with the ambiguity between scar and edema and segment them accurately. Simultaneously, the ASAN obtains the smallest HD and ASD, which means ASAN can identify the correct boundaries of scar and edema despite their indistinct shape and geometry. Moreover, we ran the statistical test on these results to measure how significantly better our results are than those from the U-Net++ and nnU-Net. As shown in Table \ref{tabpvalue2}, our method is significantly better than the other two on every metric ($p \textless 0.05$) except for the Dice of scar form U-Net++ ($p=0.109$).
Fig.
\ref{fig6} also proves that our proposed model achieves better segmentation compared to other U-Net based methods. Overall, our proposed model outperformed superiorly to other methods, demonstrating its effectiveness. This could result from the proposed deep supervision equipped with the auto-weighted mechanism to effectively extract multi-layer semantic context information. 

It is interesting to see that the vanilla U-Net scored slightly higher than the nnU-Net. Note that previous research has figured out that the nnU-Net less improves the segmentation in brain tumors than other larger targets
\citep{isensee2018nnu}. The visualized example shows that the nnU-Net method clearly underestimated the myocardial pathology region. These results suggest that the small complicated region identification might not be sensitive enough to the data augmentation and cascaded architectures used in nnU-Net. Moreover, among the three tested U-Net models, the U-Net++ method has relatively higher performance. As demonstrated in Fig.
\ref{fig6}, the U-Net++ method clearly improved performance. We conjecture that it could benefit from the connection information encoded by the dense skip connections and deep supervision. However, this U-Net++ does not consider the disparity of each layer in segmentation. On the contrary, our method utilizes an RL-based auto-weighted strategy that allows dynamic learning among layers and obtains an optimal weighting automatically.

We compared the results from the public MyoPS 2020 challenge. The corresponding results with the same Dice metric are reported in Table \ref{tab3}. Almost all teams have adopted coarse-to-fine strategies and integration strategies \citep{zhai2020myocardial,martin2020stacked,zhang2020efficientseg}. \citep{zhai2020myocardial} proposed a weighted ensemble strategy to assign different weights to the channels representing segmentation targets from the 2D and 2.5D U-Net, respectively, thus sufficiently exploiting the complementary performance of these two networks. \citep{martin2020stacked} advanced a variant of U-Net with dense convolution and bidirectional convolutional LSTM (BConvLSTM) to capture more semantic information and finally employed the averaged ensemble strategy to obtain the prediction. By contrast, a multi-task mechanism without the ensemble strategy is developed to aggregate the feature information from different modalities by employing a shared encoder and a channel reconstruction upsampling (CRU) \citep{li2020cms}. We can see that our proposed method surpasses all the competition for both scar and edema segmentation. The average Dice for scar and edema are higher than the best result in the MyoPS 2020 challenge by 0.5\% and 0.4\%, respectively. Moreover, even without using an ensemble strategy, our proposed model achieves comparable performance with the best result and outperforms other results. These findings validate the superiority of our method. Meanwhile, it is surprising that three U-Net based models also perform more superiorly than most of the competition in MyoPS 2020. The results show that the U-Net++ model tops the fourth-best Dice among all results, while the other two models obtain the sixth-best segmentation result. It might stem from the prior information used in these U-Net methods. The two-stage coarse-to-fine framework can suppress the background pixels that dominate foreground pixels in the scar and edema segmentation, thus significantly mitigate the class imbalance problem.

\subsection{The Effect of Auto-weighted Supervision}
\begin{figure}[!t]
	\centering
	\includegraphics[scale=0.44]{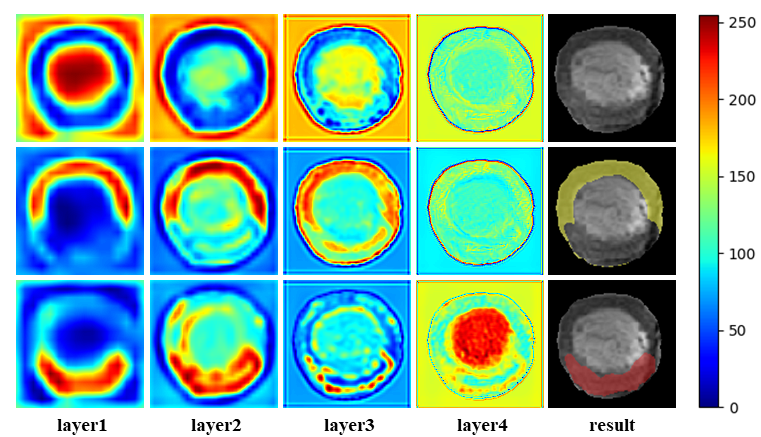}
	\caption{The visualization example of deep supervision of each layer output in scar segmentation. Among them, the first to third rows represent the background, myocardium and scar, layer 1 to 4 represent output layer from deep to shallow, respectively.}
	\label{fig7}
\end{figure}

\begin{figure}[!t]
	\centering
	\includegraphics[scale=0.44]{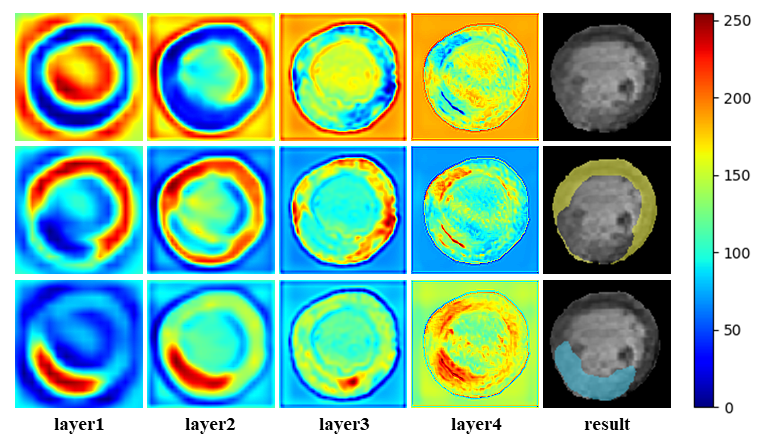}
	\caption{The visualization example of deep supervision of each layer output in edema segmentation. Among them, the first to third rows represent the background, myocardium and edema, layer 1 to 4 represent output layer from deep to shallow, respectively.}
	\label{fig8}
\end{figure}

The proposed auto-weighted module assigned [0.156, 0.200, 0.240, 0.404] to each supervised layer for the scar segmentation, while the weights from the upper to the last supervised layer were learned to be [0.217, 0.297, 0.204, 0.283] for the edema segmentation. The last supervision layer in scar segmentation gains the highest weight, highlighting its significance in making the correct identification. By contrast, the second supervised layer is almost the same important in the edema segmentation. Fig.
\ref{fig7} and \ref{fig8} 
 provide a visualization example of the feature output corresponding to each layer, including hidden layers. It shows that the upper layer represents more easily superior determinativeness, being obviously less difficult to train. On the contrary, the representation in the later layer captures more fine-grained details, thus tending to gain more predilections during training. These suggest that the smaller weight means that the corresponding layer contributes less to the gradient during training.

\begin{table*}[!t]
	\small
	\setlength{\belowcaptionskip}{4pt}%增大表格与标题的距离
	\caption{\label{tab4}The performance of training with equal weights, fixed optimal weights and automatic search weights between the supervision layers in the MyoPS dataset.}
	\centering
	\renewcommand\arraystretch{1.5}
	\resizebox{\textwidth}{!}{
		\begin{tabular}{c|cccc|c|cccc}
			\hline
			\multirow{2}{*}{Coefficients}&\multicolumn{4}{|c|}{Scar}&\multirow{2}{*}{Coefficients}&\multicolumn{4}{|c}{Scar + Edema}\\
			\cline{2-5}
			\cline{7-10}
			&Dice $\uparrow$&JC $\uparrow$&HD (mm) $\downarrow$&ASD (mm) $\downarrow$&&Dice $\uparrow$&JC $\uparrow$&HD (mm) $\downarrow$&ASD (mm) $\downarrow$\\
			\hline
			0.250,0.250,0.250,0.250 & 0.639±0.259 & 0.512±0.243 &18.07±14.48 & 2.41±3.82&0.250,0.250,0.250,0.250 & 0.705±0.122 & 0.556±0.140 & 17.86±11.77 & 1.53±2.08 \\
			0.156,0.200,0.240,0.404 & 0.646±0.253 & 0.518±0.236 & 13.19±11.11 & 1.69±3.29&0.217,0.297,0.204,0.283
			& 0.715±0.114 & 0.568±0.135 & 16.12±11.02 &1.56±1.81 \\
			Auto-weighting  &\textbf{0.658±0.253} & \textbf{0.535±0.240} & \textbf{14.13±11.90} &\textbf{1.52±2.28}&Auto-weighting& \textbf{0.720±0.126} &\textbf{0.577±0.148} & \textbf{14.12±9.24} & \textbf{1.37±2.15} \\
			\hline
	\end{tabular}}
\end{table*}

Furthermore, we examined the effect of the dynamic learning process compared to fixed-weighting. Besides the fixed-weighting with [0.250, 0.250, 0.250, 0.250] used in the above ablation experiments, we conducted another experiment with the same model and hyper-parameters, which employs the optimal set of weights from the proposed auto-weighted supervision method. Table
\ref{tab4}
 reports the performance trained with equal-weighting or fixed optimal weighting among supervision layers. The fixed optimal weighting model has relatively higher accuracy, validating our previous assumption that the supervision layer should be task-specific optimized. However, the model given optimal weighting failed to achieve the same metric as the auto-weighted approach. We contemplate that this surprising finding is owed to the RL optimization process. The RL agent continuously interacts with the environment to access the different combinations to explore the best representation from each layer freely. Meanwhile, under the RL framework, the verified information will be back-propagated to the agent as a reward signal to allow the model to learn the characteristics of a specific task better. Therefore, this kind of RL process helps to commit the entire framework to converge to a better local minimum. This finding might benefit future multi-task research by replacing fixed weights with dynamic learning methods.

We also attempted two other dynamic learning methods, including genetic algorithm and gradient propagation optimization, to search the optimal weight set of the deep supervision. The genetic algorithm takes all individuals in a group as objects and uses randomization technology to search coded parameter space efficiently. By contrast, the gradient propagation optimization updates the deep supervision weights as part of the network parameters. Also, we explored the necessity of reinforcement learning in the proposed DAS module. We still sampled $K$ times from the parameter space to generate $K$ identical networks with different weights in each epoch. We then updated the weights without using reinforcement learning by choosing the network with the best segmentation accuracy on the validation dataset (No-RL). As illustrated in Table \ref{tabgenetic}, both the genetic algorithm and gradient propagation optimization have less segmentation accuracy than our proposed method. We conjecture that the genetic algorithm only randomizes an efficient search of a coded parameter space without image-specific information, restricting its performance. On the other hand, the gradient propagation method updates the deep supervision weights through backpropagation without exploration compared to reinforcement learning. Moreover, it is interesting that there is almost no performance improvement when not using reinforcement learning to explore the optimal weights. These results suggest the image-related rewards in reinforcement learning mainly contribute to performance improvements.

\begin{table*}[!t]
	\small
	\setlength{\belowcaptionskip}{4pt}
	\caption{\label{tabgenetic}Results of different methods of finding deep supervision weights in the MyoPS dataset.}
	\centering
	\renewcommand\arraystretch{1.5}
	\resizebox{\textwidth}{!}{
		\begin{tabular}{c|cccc|cccc}
			\hline
			\multirow{2}{*}{Method}&\multicolumn{4}{|c|}{Scar}&\multicolumn{4}{|c}{Scar + Edema}\\
			\cline{2-9}
			&Dice $\uparrow$&JC $\uparrow$&HD (mm) $\downarrow$&ASD (mm) $\downarrow$&Dice $\uparrow$&JC $\uparrow$&HD (mm) $\downarrow$&ASD (mm) $\downarrow$\\
			\hline
			Genetic & 0.638±0.273 & 0.515±0.248 &18.94±14.64 & 2.46±4.80 & 0.697±0.104 & 0.549±0.149 & 19.49±12.23 & 2.01±3.02 \\
			Gradient & 0.620±0.268 & 0.495±0.246 & 20.76±14.67 & 2.40±2.97
			& 0.710±0.118 & 0.562±0.135 & 20.93±12.77 &1.95±2.02 \\
			No-RL & 0.628±0.267 & 0.502±0.243 & 22.36±15.62 & 2.99±4.73 & 0.703±0.129 & 0.556±0.147 &20.31±12.44 & 2.28±3.33\\
			Ours  &\textbf{0.658±0.253} & \textbf{0.535±0.240} & \textbf{14.13±11.90} &\textbf{1.52±2.28}& \textbf{0.720±0.126} &\textbf{0.577±0.148} & \textbf{14.12±9.24} & \textbf{1.37±2.15} \\
			\hline
			
	\end{tabular}}
\end{table*}

\begin{table*}[!t]
	\small
	\setlength{\belowcaptionskip}{4pt}
	\setlength{\tabcolsep}{4mm}
	\setlength{\belowcaptionskip}{4pt}
	\caption{\label{tabACDC}The quantitative results (Dice) of ablation experiments in the ACDC challenge. All results are from the official public evaluation platform \href{https://acdc.creatis.insa-lyon.fr/}{https://acdc.creatis.insa-lyon.fr/} .}
	\centering
	\renewcommand\arraystretch{1.5}
	\begin{tabular}{c|cc|cc|cc}
		\hline
		\multirow{2}{*}{Method}&\multicolumn{2}{|c|}{RV}&\multicolumn{2}{|c}{LV}&\multicolumn{2}{|c}{Myo}\\
		\cline{2-7}
		&ED&ES&ED&ES&ED&ES\\
		\hline
		Team1 \citep{isensee2017automatic} &\textbf{ 0.967}&\textbf{0.928}&\textbf{0.946}&\textbf{0.904}&\textbf{0.896}&\textbf{0.919} \\
		Team2 \citep{zotti2018convolutional}& 0.964&0.912&0.934&0.885&0.886&0.902 \\
		Team3 \citep{painchaud2019cardiac}& 0.961&0.911&0.933&0.884&0.881&0.897\\
		Base &0.963&0.923&0.936&0.886&0.893&0.909\\
		Base+DS&0.963&0.919&0.940&0.881&0.893&0.908\\
		Base+PAM&0.965&0.926&0.941&0.888&0.894&0.910\\
		Base+DAS&0.965&0.920&0.941&0.890&\textbf{ 0.896}&0.909\\
		Base+PAM+DAS &0.965 & 0.926 & 0.944 &0.892&\textbf{ 0.896} &0.912\\
		\hline
	\end{tabular}
\end{table*}
\begin{figure}[!t]
	\centering
	\includegraphics[scale=0.45]{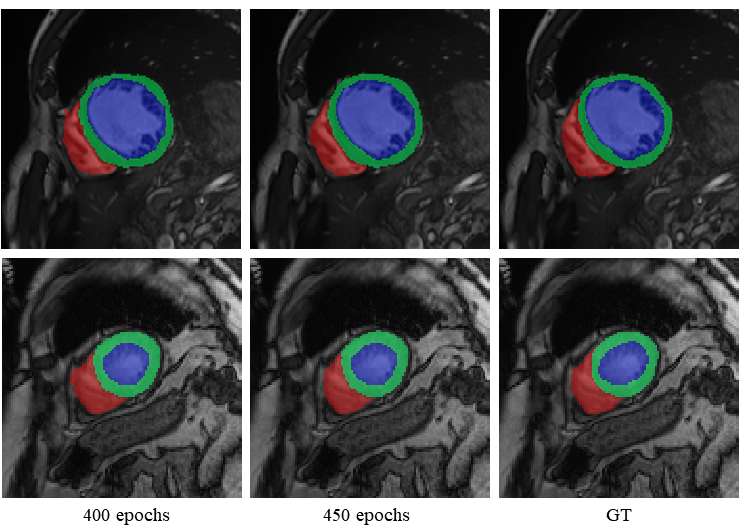}
	\caption{The segmentation visualization of different training epochs in the ACDC Challenge. Among them, RV, Myo, and LV are represented by red, green and blue respectively.}
	\label{figacdc}
\end{figure}
\begin{table*}[!t]
	\small
	\setlength{\belowcaptionskip}{4pt}
	\setlength{\tabcolsep}{5mm}
	\caption{\label{tabBUSI}The quantitative results of ablation experiments in the BUSI dataset.}
	\centering
	\renewcommand\arraystretch{1.5}
	\begin{tabular}{c|cccc}
		\hline
		&Dice $\uparrow$&JC $\uparrow$&HD (pixel) $\downarrow$&ASD (pixel) $\downarrow$\\
		\hline
		Base &0.722±0.313&0.642±0.310&39.34±45.77&12.97±20.22\\
		Base+PAM&0.732±0.317&0.657±0.310&35.68±42.25&10.87±17.88\\
		Base+PAM+DS&0.740±0.302&0.658±0.299&40.03±48.13&15.91±28.48\\
		Base+PAM+DAS &\textbf{0.747±0.290} & \textbf{0.661±0.290} & \textbf{33.81±40.20} &\textbf{9.42±20.75}\\
		\hline
	\end{tabular}
\end{table*}
\begin{figure}[!t]
	\centering
	\includegraphics[scale=0.31]{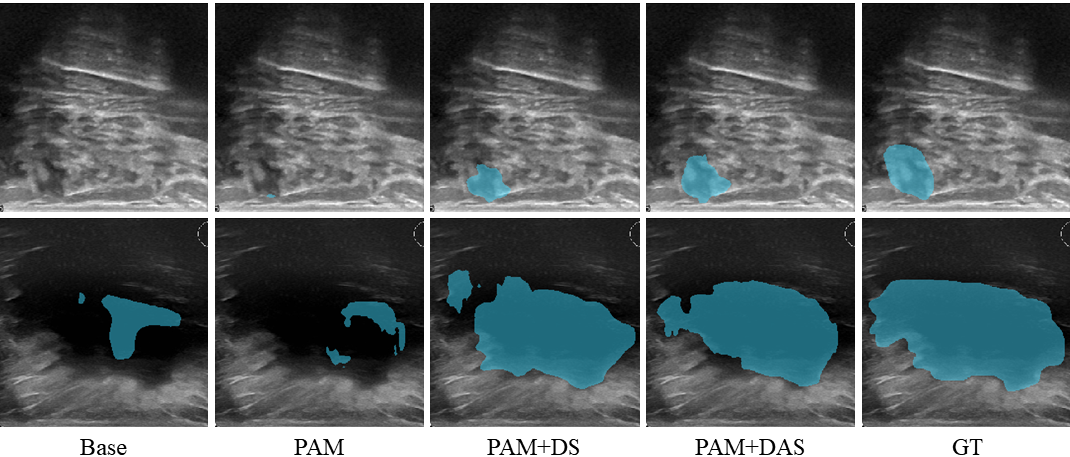}
	\caption{Visualize the segmentation results by using different training combinations in the BUSI dataset.}
	\label{figBUSI}
\end{figure}

\subsection{The Generalizability of Auto-weighted Supervision Attention Network}
We further scrutinize our proposed AWSnet on two other larger publicly available challenging datasets from different medical imaging modalities considering the limited dataset size of MyoPS 2020. The first one is a publicly available dataset from Automated Cardiac Diagnosis Challenge (ACDC) \citep{2018Deep}, supported by the MICCAI. The ACDC dataset consists of short-axis cine-MRI scans from 150 patients, including the end-diastolic (ED) and end-systolic (ES) stages. Three regions of interest, the right ventricle (RV) cavity, the myocardium (Myo), and the left ventricle (LV) cavity, delineated as the segmentation ground truth, are provided only for the 100 training cases. The second dataset is the breast ultrasound image dataset (BUSI) of 780 images from 600 female subjects. We removed 133 cases of normal subjects from the dataset to form a benchmark dataset regarding breast lesion segmentation. We follow the dataset usage in \citep{al2020dataset} to use 80 patients for training, 20 for validation, and 50 for testing. By contrast, we split the BUSI ultrasound dataset into 60\% training, 20\% validation, and 20\% test, randomly. We used the same four-layer U-Net with the same hyper-parameters in this experiment for consistency and then gradually merge each of the proposed sub-modules into the baseline model for ablation comparison implemented in a machine with an NVIDIA TITAN X(PASCAL) GPU. We trained all models for 450 epochs on the ACDC datasets and 300 epochs on BUSI datasets.

Table \ref{tabACDC} presents the segmentation results of different strategies employed on the ACDC dataset. All the results are obtained from the official public evaluation platform. Both PAM and DAS modules contributed to the performance improvement while using the deep supervision alone failed to improve the segmentation for the CMR scans in the ES stage. On the other hand, combining the PAM with the DAS module results in an obvious improvement of at least 0.2\% in all regions of interest and a maximum improvement of 1.2\% Dice for the LV cavity. We noticed that the baseline model has a satisfactory performance by a mean of above 90\% Dice. These results imply that the proposed module could be plug-in applied to any model even with a high segmentation accuracy. Fig.\ref{figacdc} shows visual comparison results at 400 epochs and 450 epochs from our method on the test set. The DAS modules contribute to fewer false positives and false negatives, although the baseline model has achieved a good segmentation accuracy. Moreover, we compared the Dice results from the public ACDC challenge. As illustrated in Table \ref{tabACDC}, our method achieves comparable performance with the best results in the official website of the ACDC challenge, suggesting the effectiveness of our proposed method.

As shown in Table \ref{tabBUSI}, our proposed model is evaluated using the Dice, JC, HD, and ASD for the BUSI dataset. Compared with the baseline model, each sub-module contributes to performance improvement in segmenting breast lesions. Using PAM brings a satisfactory segmentation performance improvement in the average Dice scores, JC, HD, and ASD. Moreover, the DAS module further boosts the performance. Overall, it can achieve total improvements of about 2 percent in Dice and JC and above 5 pixels in HDB. Fig.\ref{figBUSI} presents an example for illustrating the segmentation results from the mentioned ablation study. Our method conquers the poor image quality and boundary ambiguities and presents satisfying segmentation. These results suggest that our proposed approach has a good ability to separate each lesion pixel from the background correctly. Therefore, this kind of auto-weighted supervision helps to commit the model to focus on meaningful regions and layers, improving segmentation accuracy as an easy usage plug-in module.

\section{Conclusion}
\label{sec6}
This paper developed a fully automatic framework to segment scar and edema that are tiny and irregular targets from multi-sequence CMR images. Our framework first provides a novel design to optimize a task-specific objective trained under a reinforcement learning framework to encourage interactions among different supervised layers. We propose a two-stage coarse-to-fine framework to alleviate the bias predictions towards the majority class due to imbalanced data distribution in segmenting small structures. The cascaded coarse segmentation framework identifies the region containing edema and scar. In contrast, the fine segmentation framework extracts the myocardial pathology region on the input of LGE and T2-weighted CMR sequences cropped from the coarse network output prediction. Besides, a pixel-wise attention strategy is proposed to force the model to locate the meaningful pathological structure. The experimental results show that our method outperforms other state-of-the-art algorithms and fix-weighting deep supervision models. In the future, we will extend the proposed framework to more general multi-modality segmentation tasks. The future study will also involve more effort on applying the RL-based auto-weighted method to other multi-task applications.

\section*{Acknowledgments}
This work was supported by the Fundamental Research Funds for the Central Universities, the National Natural Science Foundation of China (NSFC 61771130 and NSFC 61801296), the National Key R\&D Program of China (2018YFA0704102) and the Shenzhen Basic Research (JCYJ20190808115419619).

%%Harvard
\bibliographystyle{model2-names.bst}\biboptions{authoryear}
\bibliography{refs}

%\section*{Supplementary Material}

%Supplementary material that may be helpful in the review process should
%be prepared and provided as a separate electronic file. That file can
%then be transformed into PDF format and submitted along with the
%manuscript and graphic files to the appropriate editorial office.

\end{document}